\documentclass[12pt,a4paper]{article}
\usepackage[utf8]{inputenc}

\usepackage{jheppub}

\usepackage{amsmath}
\usepackage[compat=1.1.0]{tikz-feynman}
\definecolor{airforceblue}{rgb}{0.36, 0.54, 0.66}
\definecolor{blue(ncs)}{rgb}{0.0, 0.53, 0.74}
\definecolor{caribbeangreen}{rgb}{0.0, 0.8, 0.6}
\definecolor{emerald}{rgb}{0.31, 0.78, 0.47}

\newcommand{\Andres}[1]{\textcolor{emerald}{#1}}
\newcommand{\Ingrid}[1]{\textcolor{magenta}{#1}}
\newcommand{\meqref}[1]{eq.~(\ref{#1})}

\def\phib{\bar{\phi}}
\def\Lmin{\mathcal{L}_\text{min}}
\def\Lnonmin{\mathcal{L}_\text{non-min}}
\def\LEM{\mathcal{L}_\text{EM}}
\def\identity{1\kern-0.25em\text{l}}
\def\S{\mathbb S}
\def\nn{\nonumber}
\def\phib{\bar{\phi}}
\def\cc{\text{c.c}}
\def\kk{q}

\def\qt{Q}

\def\sc#1{\overline{#1}}

\usepackage[normalem]{ulem}

\title{Bootstrapping Classical Spinning Compton Amplitudes with Colour-Kinematics}
\author{Ingrid Vazquez-Holm and Andres Luna}
\date{November 2022}

\begin{document}

\begin{abstract}
    {We set up a procedure to systematically obtain Compton-like amplitudes in an arbitrary-spin theory, exploiting their factorization properties, and colour-kinematics duality. We furthermore investigate the constraining of Wilson coefficients for arbitrary spinning bodies and its relation to colour-kinematic duality.}
\end{abstract}

\maketitle

\section{Introduction}
The detection of gravitational waves by the LIGO/Virgo collaboration~\cite{LIGOScientific:2016aoc,  LIGOScientific:2017vwq} kick-started a new era in multi-messenger astronomy, driving a demand for accuracy in predictions for the dynamics of the binary system in general relativity (GR). 
In this context, the connection between scattering amplitudes in quantum field theory and the two-body problem in GR~\cite{Damour:2017zjx,Bjerrum-Bohr:2018xdl} has seen increased interest. Indeed new methods developed recently allow for the extraction of potentials and physical observables from
scattering amplitudes~\cite{Cheung:2018wkq, Kosower:2018adc,
Bjerrum-Bohr:2019kec,   Brandhuber:2021eyq}. These methods aim to use modern technology developed for amplitudes, including the double copy~\cite{  Bern:2019prr} and advanced integration techniques~\cite{Parra-Martinez:2020dzs}.   
By manifestly maintaining Lorentz invariance, these amplitudes-based approaches fit naturally in the post-Minkowskian (PM) framework, where observables are expanded  in Newton's constant $G$ while keeping their exact velocity dependence.
The translation between scattering amplitudes and classical physics was used to produce the first conservative two-body Hamiltonian at $\mathcal O(G^3)$ and $\mathcal O(G^4)$~\cite{Bern:2019nnu,  Bern:2021dqo} (see also Refs.~\cite{  Dlapa:2021npj,Bjerrum-Bohr:2022ows,Jakobsen:2023ndj,Jakobsen:2023hig} for results at these orders from alternative formulations). 

New generations of gravitational-wave detectors will be much more sensitive~\cite{Punturo:2010zz,  LISA:2017pwj}, and the
spin of the black holes or neutron stars in the binary will play a more important role in the interpretation of signals.  
Furthermore, adding spin  leads to more complex dynamics, since angular momentum can be exchanged between the bodies and the orbital motion, and the system is no longer confined to a plane. This has led to a swift development in the study of the dynamics of spinning objects interacting gravitationally within the PM approximation, as a result of the application of a variety of techniques including classical GR ~\cite{Vines:2017hyw,Bini:2018ywr} (see also Refs.~\cite{Damgaard:2022jem,Hoogeveen:2023bqa,Bautista:2024agp}). 
Later, a connection between Kerr black holes and scattering amplitudes for massive spinning fields was understood \cite{Guevara:2018wpp,Chung:2018kqs}, allowing to obtain the dynamics of spinning binaries from amplitudes.

There are several ways to translate amplitudes into a description of the dynamics of the binary. 
Following the success of the non-spinning case, one of the first methods developed was the derivation of Hamiltonians for the two-body system, from which one can derive equations of motion, and in turn observables~\cite{Chung:2018kqs,Bern:2020buy,Kosmopoulos:2021zoq,Chen:2021kxt,Bern:2022kto,Alaverdian:2024spu,FebresCordero:2022jts,Gatica:2024mur}. 
Alternatively, one can bypass the Hamiltonian and go directly to observables using the method of Kosower, Maybee and O'Connell (KMOC) ~\cite{Maybee:2019jus,Menezes:2022tcs,FebresCordero:2022jts,Gatica:2023iws,Bohnenblust:2024hkw,Gatica:2024mur}, 
or by using the amplitude as a generating functional, namely the eikonal phase or radial action \cite{Guevara:2018wpp,Kosmopoulos:2021zoq,Bern:2022kto,Chen:2021kxt,Bautista:2023szu,Luna:2023uwd,Akpinar:2024meg,Alaverdian:2024spu,Chen:2024mmm,Chen:2024bpf,Aoude:2021oqj,Aoude:2022trd,Aoude:2022thd,Aoude:2023vdk,Akpinar:2025bkt}.

Besides the two-body dynamics, the waveform produced by a spinning binary has been studied in Refs. \cite{Brandhuber:2023hhl,DeAngelis:2023lvf,Aoude:2023dui,Aoude:2024jxd}, while the effect of absorption has been considered in Refs. ~\cite{Saketh:2022xjb, Aoude:2023fdm,Chen:2023qzo,Aoki:2024boe,Bautista:2024emt}.
The electromagnetic case has been shown to be similar in structure~\cite{ Bern:2021xze}, with the post-Lorentzian (PL) expansion being the  analog of the gravitational PM expansions, and so it has been used as a toy model to study spin effects~\cite{Arkani-Hamed:2019ymq,Kim:2023drc, Bern:2023ity,Kim:2024grz}. 

In parallel to the program of Amplitudes, worldline-inspired effective field theories have been used with great success for high precision computations ~\cite{Liu:2021zxr,Jakobsen:2021zvh,Jakobsen:2023ndj,Jakobsen:2023hig,Haddad:2024ebn}.  
Other related topics include applications to modified theories of gravity \cite{Brandhuber:2024bnz,Falkowski:2024bgb}, obtaining spinning metrics from amplitudes \cite{Gambino:2024uge}, and the direct connection between bound and scattering information \cite{Gonzo:2024zxo}.

In several of these recent developments, a central role has been played by the Compton amplitude, meaning a scattering amplitude with two external matter lines, and two (although it can be generalised to any integer) gauge bosons or gravitons.
The interest in the Compton amplitude is twofold. On the one hand, it acts as the building block in the constructions via generalized unitarity of the loop-level amplitudes required for higher orders in the post-Minkowskian expansion. On the other hand, the Compton amplitude is an interesting object in itself, largely because of the fascinating connection between minimal coupling (as defined in terms of massive spinor helicity by Arkani-Hamed, Huang and Huang in ref. \cite{Arkani-Hamed:2017jhn}) and the Kerr black hole, which was first explored in refs. \cite{Guevara:2017csg,Guevara:2018wpp,Chung:2018kqs}.
However, the gravitational Compton amplitude derived in that way has spurious poles beyond quartic order in the spin multipole expansion (and a similar statement can be made for minimally coupled gauge theory amplitudes beyond quadratic order in spin). This relation and its apparent breakdown, have driven a quest to find scattering amplitudes which correctly describe the Kerr black hole to any order in spin, beyond linear level in $G$. 

This interesting problem has been addressed from a variety of perspectives. 
One of them was to compute the Compton amplitude using Feynman rules, derived from a Lagrangian which is the covariant form of a worldline EFT 
~\cite{Bern:2020buy,Kosmopoulos:2021zoq,Bern:2022kto,Alaverdian:2024spu,Alaverdian:2025jtw}. While manifestly local, the EFT treatment results in the need for a matching procedure to assign values for the Wilson coefficients.
Another successful application of EFT has been the use of heavy-particle effective theories \cite{Aoude:2020onz}, which directly target classical contributions
~\cite{Aoude:2021oqj,Aoude:2023vdk}. Using these formalism, results including high orders in spin have been produced \cite{Aoude:2022trd,Aoude:2022thd,Aoude:2023vdk}, though the issue of fixing the values of Wilson coefficients remains an open problem.
On the other hand, the freedom in the Wilson coefficients allows for the description of arbitrary bodies \cite{Bern:2020buy}. The Compton amplitude for arbitrary compact objects (neutron stars) was computed using Feynman rules in ref. \cite{Bern:2022kto}, and using recursion relations (BCFW) in ref. \cite{Haddad:2023ylx}. (see also ref. \cite{Chen:2022clh} for computations involving a loop level Compton amplitude).

Alternatively, the question has been approached using massive higher-spin quantum field theories, attempting to use their properties to predict the amplitude of the black hole~\cite{Chiodaroli:2021eug, Cangemi:2022abk, Cangemi:2022bew,Ochirov:2022nqz,Cangemi:2023ysz,Cangemi:2023bpe}. The problem of defining a consistent quantum field theory with higher spin is a long-storied one, but it recently has seen a lot of progress with the application of chiral massive fields \cite{Ochirov:2022nqz,Cangemi:2023ysz,Cangemi:2023bpe}.

Other approaches which rely more on the classical side include Refs.~\cite{Bautista:2022wjf,Bautista:2023sdf} where solutions of the Teukolsky equation are used to derive the classical limit of the Compton amplitude (see also Refs. ~\cite{Ben-Shahar:2023djm,Scheopner:2023rzp} for the derivation of classical amplitudes for also for neutron stars from the worldline).

Given the role of the massive spinor helicity amplitude and its connection with the Kerr black hole, the discussion has largely centered around helicity Compton amplitudes. However, it is possible to study fully covariant forms of these amplitudes. This was done in refs. \cite{Kosmopoulos:2021zoq,Bern:2022kto}, resulting in long, rather structureless expressions, as expected from their Feynman-diagram origin. More recently, compact covariant forms of the Compton amplitude have been derived using the so-called HEFT \cite{Bjerrum-Bohr:2023jau,Bjerrum-Bohr:2023iey,Bjerrum-Bohr:2024fbt},   
while another covariant very compact form of the Compton amplitude was given in ref. \cite{DeAngelis:2023lvf}, and whose properties mesh well with the double copy.

As mentioned above, one of the advantages of  using an effective quantum field theory with free Wilson coefficients to be fixed by matching an observable as in refs. \cite{Bern:2020buy,Kosmopoulos:2021zoq,Bern:2022kto,Alaverdian:2024spu}, is that the Wilson coefficients allow for the description of more generic bodies than black holes, for example neutron stars. In this context, we argue the study of the properties of Compton amplitudes with free Wilson coefficients to be important in itself. One such property that concerns us is the ability to obtain them through a bootstrap procedure, perhaps involving the double copy.

In this project, we aim to make contact with both the utilitarian approach of using them as building blocks for loop-level amplitudes, and the fundamental aspect of the Compton amplitude. Our objective is to set up a procedure to systematically obtain Compton-like amplitudes
in an arbitrary spin theory. In this paper we look at 4- and 5-point amplitudes (meaning two massive and two or tree massless), with or without spin, and with or without free Wilson coefficients.

On the field theory side, we consider effective field theories where the spin-induced multipoles are parameterized by Wilson coefficients, whose value may be fixed by matching to a full theory computation. They result in scattering amplitudes with spin structures either in terms of Lorentz generators (quantum) or spin tensors (classical).
This has been done before by considering a Lagrangian, obtained as a covariant generalisation of a worldline theory\cite{Bern:2020buy}. 
Instead, one may systematically list the quantum spin structures to make an ansatz, that we use as seed for a bootstrap procedure described below.

The prescription for bootstrapping spinning amplitudes that we introduce here closely follows the setup for scalar massive particles in \cite{Carrasco:2020ywq}. As one of the goals of this work is to probe the interplay between the color-kinematics duality of spinning amplitudes and their Wilson coefficients, we obtain the kinematic numerators of the amplitude graphs in a way that obeys the duality by construction. Incidentally, this allows us to express the numerators of the graphs in terms of a set of basis graph numerators. These basis numerators are then given the aforementioned ansatze made up of the allowed quantum spin structures. We fix the free coefficients in the ansatze by imposing that numerators obey the same symmetry constraints and factorization properties as their corresponding graphs, and by imposing that the resulting amplitudes (both in the gauge theory and gravity) obey gauge invariance.

We begin in section \ref{sec::Bootstrap} by establishing the basics of the bootstrap procedure, and we apply it first to linear-in-spin amplitudes. Once the procedure has been described, we use it to obtain quadratic-in-spin results.
In section \ref{sec::Gravity}, we use the bootstrapped gauge theory amplitude to produce results in gravity and compare to previous results in the literature. We then turn to the question of applying the bootstrap procedure at the cubic-in-spin order.
Finally, in section \ref{sec::discussion} we discuss our results and conclude.

\section{Bootstrapping Spinning Amplitudes}
\label{sec::Bootstrap}

\subsection{The three-point amplitude}

The main building block for the amplitudes we bootstrap in this paper is the three-point amplitude with massive spinning matter interacting with a gauge boson (or graviton). This amplitude is represented by the following graph, 
\begin{equation}
\mathcal{A}(p_1, p_2, q) =
\begin{gathered}
    \includegraphics[scale=1]{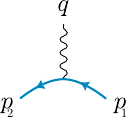}
\end{gathered},
\end{equation}
where the solid blue line represents the massive spinning particle, and the wavy line represents the gauge boson. This three-point amplitude has previously been deduced from a Lagrangian \cite{Bern:2020buy}, but here we determine it up to cubic-in-spin order from the same principles we will use to bootstrap the higher-multiplicity amplitudes in the following sections. Our scattering amplitudes will be given in terms of hermitian Lorentz generators in the representation
\begin{equation}
(M^{ab}){}_{c(s)}{}^{d(s)} = {2 {} i {} s}\delta^{[a}_{(c_1}\eta^{b](d_1}\delta^{d_2}_{c_2}\dots \delta^{d_s)}_{c_s)}\,,
\hskip 1.5 cm 
(M^{ab}){}_{c(s)}{}_{d(s)} = -(M^{ab}){}_{d(s)}{}_{c(s)} \, ,
\label{ExplicitLorentzGenerators}
\end{equation}
where the indices $c(s)$ and $d(s)$ stand for the symmetrized 
sets of vector indices $\{c_1,\dots, c_s\}$ and $\{d_1,\dots, d_s\}$, respectively.
The generators $M^{ab}$ satisfy the usual Lorentz algebra,
\begin{equation}
[M^{a_1 a_2},M^{a_3 a_4}]=i(
\eta^{a_3 a_1}M^{a_4 a_2}
+\eta^{a_2 a_3}M^{a_1 a_4}
-\eta^{a_4 a_1}M^{a_3 a_2}
-\eta^{a_2 a_4}M^{a_1 a_3}) \, ,
\label{LorentzAlgebra}
\end{equation}
where for conciseness, we have and will from hereon omit the spin indices $a(s)$, etc. in the Lorentz generators. In general we decompose the products of generators as
\begin{align}
M^{\mu_1\nu_1}M^{\mu_2\nu_2}=&
\{M^{\mu_1\nu_1},M^{\mu_2\nu_2}\}+
[M^{\mu_1\nu_1},M^{\mu_2\nu_2}],\\
M^{\mu_1\nu_1}M^{\mu_2\nu_2}M^{\mu_3\nu_3}=&
\{M^{\mu_1\nu_1},M^{\mu_2\nu_2},M^{\mu_3\nu_3}\}+
\{M^{\mu_1\nu_1},[M^{\mu_2\nu_2},M^{\mu_3\nu_3}]\}\\
\nonumber
&
+
\{M^{\mu_2\nu_2},[M^{\mu_3\nu_3},M^{\mu_1\nu_1}]\}
+
\{M^{\mu_3\nu_3},[M^{\mu_1\nu_1},M^{\mu_2\nu_2}]\}\\
\nonumber
&+\frac{2}{3}[M^{\mu_1\nu_1},[M^{\mu_2\nu_2},M^{\mu_3\nu_3}]]
-\frac{2}{3}[M^{\mu_3\nu_3},[M^{\mu_1\nu_1},M^{\mu_2\nu_2}]],
\end{align}
because it is the (weighted) totally symmetric product of Lorentz generators 
\begin{align}
\{M_{\mu_1\nu_1},M_{\mu_2\nu_2},\ldots,
M_{\mu_n\nu_n}\}\equiv \frac{1}{n!}(M_{\mu_1\nu_1}M_{\mu_2\nu_2}\ldots
M_{\mu_n\nu_n}+\textrm{perms.})
\end{align}
that results in the classical-limit relation \cite{Bern:2020buy}
\begin{align}
\label{eq:classicallimit}
    M^n(v_1,v'_1,\ldots,v_n,v'_n)
    =
    S(v_1,v_1')S(v_2,v_2')\ldots S(v_n,v_n'),
\end{align}
where $S(a,b)\equiv S^{\mu\nu}a_\mu b_{\nu}$, and $S^{\mu\nu}$ is the classical spin tensor. Furthermore we introduce the notation
\begin{align}
    M^n(v_1,v'_1,\ldots,v_n,v'_n)
    \equiv
    \{M_{\mu_1\nu_1},
    M_{\mu_2\nu_2},\ldots,
    M_{\mu_n\nu_n}\}
    {v_1}^{\mu_1}{v'_1}^{\nu_1}\ldots
    {v_n}^{\mu_n}{v'_n}^{\nu_n}.
\end{align}
Originally, the relation contains a product of the polarization tensors for the massive legs $\varepsilon^s_1 \cdot \varepsilon^s_2$, which factorizes from the generators product. The net effect of that product is a change from a so-called canonical position (impact parameter) to the covariant one that we use here (see ref. \cite{Bern:2020buy}, for details on this topic), and so we omit it.
The goal here is to determine the kinematic part of the three-point amplitude above, which can then be used to describe the amplitude in any given gauge theory given the proper charge. We express the kinematics of our amplitudes in terms of Lorentz products between momenta and Lorentz generators. We set up the following ansatz $\mathfrak{A}(p_1, p_2, q)$ for the terms that could possibly show up in a three-point amplitude consisting of these building blocks, 
\begin{multline}
    \mathfrak{A}(p_1, p_2, q) =
    w_1 ~(p_2\cdot \varepsilon_q)( \varepsilon ^s_1\cdot \varepsilon ^s_2)
    +w_2~ M(q,\varepsilon _q)
    +w_3 ~M(p_1,\varepsilon _q)
    -\frac{w_4}{m^2} ~(p_2\cdot \varepsilon _q) M(p_1,q)
    \\
-\frac{w_5}{m^2} M^2( p_1, q, q, \varepsilon_q)
+\frac{w_6}{m^4} (p_2\cdot \varepsilon _q) 
M^2(p_1, q, p_1, q)
\\
 +\frac{w_7}{m^2}~( p_2\cdot \varepsilon _q) 
 M^2( q, \xi,\xi,q)
 +\frac{w_8}{m^4} M^3( p_1, q, p_1, q, q, \varepsilon _q)
 \\
  +\frac{w_9}{m^2}
 ~M^3(\xi,
 q,\xi, q, q, \varepsilon_q)
   -\frac{w_{10}}{m^6} (p_2\cdot \varepsilon_q) 
    M^3(
    p_1, q, p_1, q, p_1, q)
    \\
    -\frac{w_{11}}{m^4} (p_2\cdot \varepsilon _q)  M^3
    (\xi, q, \xi, q, p_1, q),   
    \label{threepointseed}
\end{multline}
where $m^2$ is the mass square of the spinning particle, $\varepsilon_q$ is the polarization vector of the gauge boson, $\varepsilon_i^s$ is the polarization of the massive spinning particle $i$, and $w_i$ are the free coefficients of the ansatz.
Note that we are allowing masses in the denominators of the terms --- this could at first glance imply that the ansatz for this amplitude would be infinitely large, but it is in fact strongly constrained by the following conditions: (1) the only independent dot products after imposing conservation of momenta are $m^2$  and $\varepsilon_3 \cdot p_2$, (2) the polarization vector $\varepsilon_3$ must show up once, and only once, in each term, and (3) we allow one power of $q$ for every Lorentz generator in the ansatz, to correctly reproduce the classical limit.  

In addition to these constraints, we demand the amplitude to be gauge invariant and antisymmetric under the exchange $p_1 \leftrightarrow p_2$. We note that this anti-symmetry between massive lines is not imposed explicitly at higher-point amplitudes, as even-power products of Lorentz generators are not necessarily antisymmetric under this exchange. This is in contrast to the bootstrapping of scalar amplitudes in \cite{Carrasco:2020ywq} where gauge invariance is automatically satisfied when the amplitudes obey all symmetries, and consequently we need to impose gauge invariance in order to fix all information for the spinning amplitudes here. 
We can, however, impose anti-symmetry for the three-point amplitude, and find that $w_3=0$. The surviving $w_i$ will remain free and become our Wilson coefficients. The bootstrap procedure allows to fix the amplitude up to an overall constant, that we determine by matching to the three-point scalar. 

Previous papers based on the arbitrary spin formalism used Lagrangians, and subsequently Feynman diagrams to obtain the amplitudes. Let us, for illustration, briefly review these quantum field theories. The Lagrangian gives a covariantisation of spin-induced multipole moments, and is interpreted as an effective theory, valid only at sufficiently large impact parameter, i.e. only in the classical regime.
The Lagrangian is separated into a minimal and non-minimal part.
The minimal coupling Lagrangian for QED involves only the standard two-derivative kinetic terms,
where all the massive bodies are taken as carrying the same charge $Q$ (the PL framework expands observables in powers of $\alpha\equiv Q^2/(4\pi)$ keeping the exact velocity dependence). In ref. \cite{Bern:2023ity}, the following non-minimal linear-in-$F_{\mu\nu}$ interactions up to two powers of spins are considered,
\begin{align}\label{eq:LS1}
	&\Lnonmin 
	= Q C_1 F_{\mu\nu} \phi_s M^{\mu\nu} \phib_s + \frac{QD_1}{m^2}F_{\mu\nu} ( D_{\rho}\phi_s M^{\rho\mu}D^{\nu}\phib_s + \cc)\\
	& \quad -\frac{iQ C_2}{2m^2}\partial_{(\mu}F_{\nu)\rho}(D^{\rho}\phi_s \nn \S^{\mu}\S^{\nu}\phib_s-\cc) 
 -\frac{iQ D_2}{2m^2} \partial_{\mu}F_{\nu\rho}(D_{\alpha}\phi_s M^{\alpha\mu}M^{\nu\rho}\phib_s-\cc)\,,\nonumber
\end{align}
where $\S^{\mu}\equiv \frac{-i}{2m}\epsilon^{\mu \nu \rho \sigma}M_{\nu \rho}D_{\sigma}$ is the Pauli-Lubanski vector, and $D_{\sigma}$ is the covariant derivative. 
In general, the Wilson coefficients $C_i$ and $D_i$ of these operators need to be matched to either theoretically or experimentally determined values, with coefficients being particularly simple for black holes.
From this, one may construct the three-point amplitude. One could also have included a third Wilson coefficient at quadratic in spin order ($F_2$). This was omitted in ref. \cite{Bern:2023ity}, because it was beyond the scope of that paper, but we do include it here for completeness. To keep the notation consistent with this treatment, we rename the coefficients in the three-point amplitude eq. (\ref{threepointseed}) to match those appearing in the  Lagrangian\footnote{Notice that this differs from ref. \cite{Bern:2023ity} by a sign in front of Wilson coefficient $C_1$. This comes from lining up three-point amplitudes with different ordering, \textit{i.e.} $A(p_1, p_2, q)$ versus $A(p_2, p_1, q)$, which leads to a sign difference in the Lorentz generators $M$ from flipping $p_1 \leftrightarrow p_2$.}, 
\begin{multline}\label{Eq: 3-point amplitude}
   A(p_1, p_2, q) =
    2 (p_2\cdot \varepsilon_q)( \varepsilon ^s_1\cdot \varepsilon ^s_2)
    +2 i C_1 M(q,\varepsilon _q)
    -2 i\frac{D_1}{m^2} ~(p_2\cdot \varepsilon _q) M(p_1,q)
    \\
+2\frac{D_2}{m^2}  M^2( p_1, q, q, \varepsilon_3)
-\frac{F_2}{m^4} (p_2\cdot \varepsilon _q) 
 M^2(p_1, q,p_1, q)
\\
 +\frac{C_2}{m^2}~( p_2\cdot \varepsilon _q) M^2(q,\xi, q, \xi)
 +\frac{C_3}{m^4} M^3(p_1, q, p_1, q,q, \varepsilon _q)
 \\
  +\frac{D_3}{m^2}
 ~M^3(q, \xi, q, \xi, q, \varepsilon _q)
   -\frac{F_3}{m^6} (p_2\cdot \varepsilon _q) 
    M^3( p_1, q, p_1, q, p_1, q)
    \\
    -\frac{G_3}{m^4} (p_2\cdot \varepsilon _q)  ~M^3(
    q,\xi, q, \xi, p_1, q).  
\end{multline}
One last building block to be used in the bootstrap is the three-point amplitude in Yang-Mills theory, 
\begin{equation}\label{Eq: 3-point YM}
   A_{\text{YM}}(k_1,k_2,k_3) = (\varepsilon _1 \cdot \varepsilon _2)( k_2\cdot \varepsilon _3)+(\varepsilon _1\cdot \varepsilon _3)( k_1\cdot \varepsilon _2)+(\varepsilon _2\cdot \varepsilon _3)( k_3\cdot \varepsilon _1). 
\end{equation}
We will now bootstrap four- and five-point amplitudes using three-point amplitudes as seeds.

\subsection{Bootstrapping four-point amplitudes}
We're now ready to construct the four-point amplitude with a massive spinning particle and two emitted gauge bosons, known as the \textit{Compton amplitude}. We set up a bootstrapping scheme for the kinematic numerators of the graphs that contribute to the YM amplitudes, such that they -- by construction -- obey the color-kinematics duality. This will allow us to take the direct double-copy graph-by-graph in order to obtain gravity numerators. As the kinematic information of the gauge theory graphs is independent of the specifics of the gauge theory, we can use the same numerators to obtain the QED amplitudes.  The numerators should factorize into the correct lower-point amplitudes when propagators are put on-shell, and this will set further constraints on the bootstrapped numerators. 

Consider the four-point Compton amplitude, which can be expressed in terms of the following graph topologies, 
\begin{equation}
\label{Eq:Compton graphs}
\begin{aligned}
\begin{gathered}
    \includegraphics[scale=1.0]{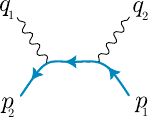}
\end{gathered}
\hspace{0.5cm}
&,
\hspace{0.5cm}
\begin{gathered}
    \includegraphics[scale=1.0]{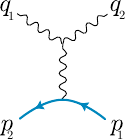}
\end{gathered},
\\
n^4_1(p_1, p_2, q_1, q_2)
\hspace{0.4cm}
&
\hspace{0.4cm}
n^4_2(p_1, p_2,q_1,q_2)
\end{aligned}
\end{equation}
where the associated kinematic numerator $n_i^4$ is stated below each topology. Note that we will only use cubic topologies -- as we will be constructing inherently color-dual numerators, any information from contact terms is distributed amongst the cubic graphs. In terms of graphs the QED and YM amplitudes will then be given as, respectively, 
\begin{equation}
\label{Eq: 4-point amplitudes def qedym}
    \begin{aligned}
\mathcal{A}^{\text{QED}}_4(p_1, p_2,q_1,q_2) &=\hspace{0.5cm}
    \begin{gathered}
\includegraphics[scale=1.0]{figures/spin_4p_t.pdf}
\end{gathered}
\hspace{0.3cm}
+
\hspace{0.5cm}
  \begin{gathered}
\includegraphics[scale=1.0]{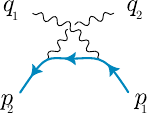}
\end{gathered}
\\
&=Q_1  \frac{n_1(p_1,p_2,q_1,q_2)}{(p_1+q_2)^2 - m^2}+Q_1 \frac{n_1(p_1,p_2,q_2,q_1)}{(p_1+q_1)^2 - m^2},
\\
\mathcal{A}^{\text{YM}}_4(p_1,p_2,q_1,q_2) &=
    \hspace{0.4cm}
    \begin{gathered}
\includegraphics[scale=1.0]{figures/spin_4p_t.pdf}
\end{gathered}
\hspace{0.4cm}
+
\hspace{0.4cm}
  \begin{gathered}
\includegraphics[scale=1.0]{figures/spin_4p_u.pdf}
\end{gathered}
\hspace{0.5cm}
+
\hspace{0.2cm}
\begin{gathered}
\includegraphics[scale=1.0]{figures/spin_4p_s.pdf}
\end{gathered}
\\
&= c_1 \frac{n_1(p_1,p_2,q_1,q_2)}{(p_1+q_2)^2 - m^2}+c_2 \frac{n_1(p_1,p_2,q_2,q_1)}{(p_1+q_1)^2 - m^2}+c_3 \frac{n_2(p_1,p_2,q_1,q_2)}{(p_1+p_2)^2},
 \end{aligned}
\end{equation}
where $Q_i$ are QED charges and $c_i$ are YM color factors. The gravity amplitude is therefore given by
\begin{equation}
\label{Eq: 4-point amplitudes def gr}
\begin{aligned}
    \mathcal{M}_4(p_1,p_2,q_1,q_2) 
    &=\hspace{0.2cm}
    \begin{gathered}
\includegraphics[scale=1.0]{figures/spin_4p_t.pdf}
\end{gathered}
\hspace{0.2cm}
+
\hspace{0.3cm}
  \begin{gathered}
\includegraphics[scale=1.0]{figures/spin_4p_u.pdf}
\end{gathered}
\hspace{0.3cm}
+
\hspace{0.2cm}
\begin{gathered}
\includegraphics[scale=1.0]{figures/spin_4p_s.pdf}
\end{gathered}
\\
\nonumber 
&=\frac{{n_1(p_1,p_2,q_1,q_2)}^2}{(p_1+q_2)^2 - m^2}+\frac{{n_1(p_1,p_2,q_2,q_1)}^2}{(p_1+q_1)^2 - m^2}+\frac{{n_2(p_1,p_2,q_1,q_2)}^2}{(p_1+p_2)^2}.
    \end{aligned}
\end{equation}
We further demand that the kinematic numerators of the graphs obey the same Jacobi relation as the color factors, 
\begin{equation}\label{Eq: 4-point Jacobi}
\begin{aligned}
    \begin{gathered}
\includegraphics[scale=1.0]{figures/spin_4p_s.pdf}
\end{gathered}
=&
\hspace{0.2cm}
 \begin{gathered}
\includegraphics[scale=1.0]{figures/spin_4p_t.pdf}
\end{gathered}
\hspace{0.1cm}
-
\hspace{0.1cm}
 \begin{gathered}
\includegraphics[scale=1.0]{figures/spin_4p_u.pdf}
\end{gathered},\\
n_2(p_1,p_2,q_1,q_2)~=&~ n_1(p_1,p_2,q_1,q_2)-n_1(p_1,p_2,q_2,q_1),
\end{aligned}
\end{equation}
from which we see that $n_2$ can be obtained as a linear combination of $n_1$. We will therefore give an ansatz only to $n_1$, which we will call the \textit{basis numerator}. 

\subsubsection{The ansatz}
The numerator ansatz should consist of all the kinematic terms we can write down in a minimal kinematic basis, while ensuring every term respects power counting. In order to set up expressions we take a closer look at the graph for $n_1$, given in \meqref{Eq:Compton graphs}. 
The kinematics will consist of combinations of (1) Lorentz products $k_i \cdot k_j$, and (2) spin matrices $M(k_i, k_j)$. Conservation of momenta allows us to remove one of the momenta,
\begin{equation}
    p_2 \rightarrow -p_1-q_1-q_2,
\end{equation}
therefore a minimal basis of Lorentz products between momenta and polarizations is then,
\begin{equation}
  \beta^L_i \in  \{ p_1 \cdot p_2~,~ p_1 \cdot q_2~, ~m^2~, ~p_1 \cdot \varepsilon_1~,~ p_1 \cdot \varepsilon_2~,~ q_2 \cdot \varepsilon_1~,~ q_1 \cdot \varepsilon_2\},
\end{equation}
where $m^2$ is the square mass of the massive spinning particle, $\varepsilon_i$ is the polarization of the massless particle $q_i$, and $\varepsilon^s_i$ is the polarization of the massive spinning particle $p_i$. 
We will eventually take the classical limit 
\meqref{eq:classicallimit}, 
and furthermore impose the spin supplementary condition (SSC), ${p_1}_{\mu} S^{\mu\nu} = 0$, so we can discard terms of the form $M(p_1,k_i)$. Thus, the basis of spin matrices contains the following elements,
\begin{equation}
  \beta^M_i \in  \{ M(q_1, q_2)~,~ M(q_1, \varepsilon_1)~,~ M(q_2, \varepsilon_1)~,~ M(q_1, \varepsilon_2)~,~ M(q_2, \varepsilon_2)\}.
\end{equation}
and the most general ansatz, up to linear-in-spin order we can write down is then\footnote{This is the most general ansatz considering the coupling constants in the three-point amplitude contain at most $m^4$ in the denominator.}, 
\begin{equation}
\label{Eq:4-point linear ansatz}
\begin{aligned}
\mathfrak{A}_1(p_1,p_2,q_1,q_2) =& ~\sum_l a_l \beta^L_i \beta^L_j (\varepsilon^s_1 \cdot \varepsilon^s_2)
+\sum_m b_m \beta_i^L \beta_j^M
+\sum_n c_n\beta_i^L \beta_j^L \beta_k^M
\end{aligned}
\end{equation}
where  $a_l$, $b_m$ and $c_n$ are free coefficients for the 132 total terms that appear in the ansatz.

\subsubsection{Factorization channels}

The next step in our bootstrap is to calculate the factorization channels of the amplitude --- it should factorize into known three-point amplitudes when propagators are put on-shell. The factorization channels allow us not only to fix the ansatz coefficients, but also investigate the interplay between Wilson coefficients and the color-kinematics duality. Since there are two topologies, there are two factorization channels, and the three-point amplitudes are sewn together using $D$-dimensional state projectors. The well-known gluon projector is,
\begin{equation}
\varepsilon^{\mu}(k) \varepsilon^{\nu}(k) \equiv \eta^{\mu \nu} - \frac{k^{\mu} q^{\nu} + q^{\mu} k^{\nu}}{k \cdot q},
\end{equation}
where $q$ is a reference vector. We choose the projector for massive spinning states as, 
\begin{equation}  \varepsilon^s(k)^{a(s)}\varepsilon^s(-k)^{b(s)} \to \eta^{a(s)b(s)}.
\end{equation}
As discussed extensively elsewhere \cite{Bern:2023ity}, this has the effect of not imposing transversality (thus allowing different degrees of freedom to propagate). Eventually, this is equivalent to fixing Wilson coefficients to specific values. 

For the purpose of investigating the color-kinematics duality we set up a discrepancy function $\Delta$. This function will allow us to verify whether the duality is in fact consistent with the factorizations of the amplitude, up to terms proportional to inverse propagators\footnote{These will be set to zero by the cut conditions in some factorizations but not others.},
\begin{equation}\label{Eq: pseudo-Jacobi}
\Delta
=
    \begin{gathered}
\includegraphics[scale=1.0]{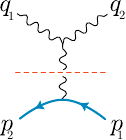}
\end{gathered} 
-
 \begin{gathered}
\includegraphics[scale=1.0]{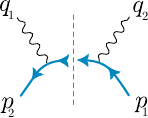}
\end{gathered}
+
\begin{gathered}
\includegraphics[scale=1.0]{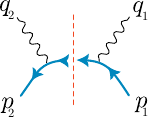}
\end{gathered}
+
\text{inv. prop.},
\end{equation}
where we expect $\Delta = 0$ for a color-kinematics dual representation. 

We now take a closer look at the factorizations. The massless 
factorization is given by sewing together the three-point amplitudes in \meqref{Eq: 3-point amplitude} and \meqref{Eq: 3-point YM},
\begin{equation}\label{Eq: 4-point linear massless factorization}
\begin{aligned}
           \begin{gathered}
\includegraphics[scale=1.0]{figures/spin_4p_s_cut.pdf}
\end{gathered} 
= 
& \sum_{\text{states}} A(p_1, p_2, l)A_{\text{YM}}(-l, q_1, q_2)
\\
 = &~  -4 i C_1 \Big(
       2 (p_1\cdot \varepsilon _2) M(q_{12},\varepsilon _1)-2 (p_1\cdot \varepsilon _1 )M(q_{12},\varepsilon _2)\\
       &\hspace{6cm}-(\varepsilon _1\cdot \varepsilon _2) M(q_1,q_2)
       \Big),
        \end{aligned}
    \end{equation}
where $q_{12}=q_1+q_2$. Note that only the $C_1$ Wilson term survives here.
Similarly, the massive factorization is given by,
\begin{equation}\label{Eq: mass factorization}
\begin{aligned}
        \begin{gathered}
\includegraphics[scale=1.0]{figures/spin_4p_t_cut.pdf}
\end{gathered}
=&  \sum_{\text{states}} A(p_2, q_1, l)
A(-l, q_2, p_1)
\\
=&  ~\mathfrak{C}_{\beta_1}+ \mathfrak{C}_{\beta_2},
\end{aligned}
\end{equation}
where we have split the cut into a term $\mathfrak{C}_{\beta_1}(\beta^M_1)$ in terms of a basis of $M$'s that show up both in the massless and massive factorization channels: 
\begin{equation}
\beta^M_1 = 
\{M(q_1, \varepsilon_1) , M(q_1, \varepsilon_2) , M(q_2, \varepsilon_1) , M(q_2, \varepsilon_2) , M(q_1, q_2)\},
\end{equation}
and a term $\mathfrak{C}_{\beta_2}(\beta^M_2)$ in terms of a basis on $M$'s that \textit{only} appear in the massive factorization channel, 
\begin{equation}
  \beta_2^M= \{ M(\varepsilon_1, \varepsilon_2)\}.
\end{equation} 
Explicitly, the two factorization terms are,
\begin{align}
    \mathfrak{C}_{\beta_1}
    =& ~
2 i  \left( C_1^2 \varepsilon _1\cdot \varepsilon _2-\frac{ D_1 }{m^2}p_2\cdot \varepsilon _1 \left(\left(D_1+C_1-2\right) p_1\cdot \varepsilon_2 - C_1 p_2\cdot \varepsilon _2\right)\right) M(q_1,q_2)
  \nonumber
  \\
  &+2 i C_1  \left(C_1 p_{12}\cdot \varepsilon _1+\frac{D_1}{m^2} p_{12}^2 p_2\cdot \varepsilon _1\right) M(q_1,\varepsilon _2) -2 i C_1^2 p_{12}\cdot \varepsilon _2 M(q_2,\varepsilon _1)
  \nonumber
  \\
  &+2 i C_1 \left(\frac{D_1}{m^2} p_{12}^2-2 \right) p_1\cdot \varepsilon _2 M(q_1,\varepsilon _1)
 +4 i C_1 p_2\cdot \varepsilon _1 M(q_2,\varepsilon _2),
  \\
  \mathfrak{C}_{\beta_2}
        =&~ 2 i C_1^2 (p_1+p_2)^2 M(\varepsilon _1,\varepsilon _2),
\end{align}
where $p_{12}=p_1+p_2$. The other massive factorization channel, the third term in \meqref{Eq: pseudo-Jacobi}, is simply a relabeling $q_1 \leftrightarrow q_2$ of these expressions.
The difference between the two massive factorizations can also be split into two terms, $\Delta\mathfrak{C}_{\beta_1}$ and $\Delta \mathfrak{C}_{\beta_2}$, in the two separate bases of $M$'s, $\beta_1^M$ and $\beta_2^M$,
\begin{equation}
     \begin{gathered}
\includegraphics[scale=1.0]{figures/spin_4p_t_cut.pdf}
\end{gathered}
-
\begin{gathered}
\includegraphics[scale=1.0]{figures/spin_4p_u_cut.pdf}
\end{gathered}
= \Delta \mathfrak{C}_{\beta_1} + \Delta \mathfrak{C}_{\beta_2}.
\end{equation}
From the definition of the discrepancy function $\Delta\mathfrak{C}_{\beta_2}$ should vanish (up to inverse propagators) when taking the difference between the two massive factorizations, as there are no $\beta_2^M$-terms in the massless factorization. Similarly, $\Delta\mathfrak{C}_{\beta_1}$ should cancel the massless factorization up to inverse propagators.
Relabeling and taking the difference between massive factorizations we find,
\begin{equation}
    \Delta \mathfrak{C}_{\beta_2} =  \mathfrak{C}_{\beta_2} - \mathfrak{C}_{\beta_2} \big|_{q_1 \leftrightarrow q_2} = 4 i C_1^2 (p_1+p_2)^2 M(\varepsilon _1,\varepsilon _2) ,
\end{equation}
where, as $(p_1+p_2)^2$ is the inverse massless propagator which is set to zero in the massless factorization, we find that this term satisfies our conditions. 
The term $\Delta \mathfrak{C}_{\beta_1}$ is slightly more involved,
\begin{equation}
\label{Eq: 4-point linear massive diff}
\begin{aligned}
     \Delta \mathfrak{C}_{\beta_1} =& ~ \mathfrak{C}_{\beta_1} - \mathfrak{C}_{\beta_1} \Big|_{q_1 \leftrightarrow q_2}\\
=& \hspace{0.2cm}
4 i C_1^2(\varepsilon _1\cdot \varepsilon _2) M(q_1,q_2)\\
&+
2 i \frac{D_1}{m^2} \Big[(p_2\cdot \varepsilon _2 )\left(W_1 (p_1\cdot \varepsilon _1)+2 C_1 (p_2\cdot \varepsilon _1)\right)+W_1 (p_1\cdot \varepsilon _2)( p_2\cdot \varepsilon _1)
\Big]M(q_1,q_2)\\
&
-
4 i C_1 \Big[(
p_{12}\cdot \varepsilon _1) \left(-C_1 M(q_1,\varepsilon _2)-M(q_2,\varepsilon _2)\right)
+p_{12}\cdot \varepsilon _2 \left(M(q_1,\varepsilon _1)+C_1 M(q_2,\varepsilon _1)\right)
\Big]\\
&+ (p_1+p_2)^2 \Delta \mathfrak{C}_s
,
\end{aligned}
\end{equation}
where we define the combination of Wilson coefficients $W_1\equiv 2-C_1-D_1$ and  $\Delta \mathfrak{C}_s$ is the coefficient of the inverse propagator $(p_1+p_2)^2$, which is irrelevant for the purpose of this analysis. Comparing with \meqref{Eq: 4-point linear massless factorization} we immediately see that the second line  of \meqref{Eq: 4-point linear massive diff} requires $D_1 = 0$. The remaining terms are, 
\begin{align}
\Delta \mathfrak{C}_{\beta_1}
=&
    -4 i C_1 \Big(p_{12}\cdot \varepsilon _1 \left(-C_1 M(q_1,\varepsilon _2)-M(q_2,\varepsilon _2)\right)+p_{12}\cdot \varepsilon _2 \left(M(q_1,\varepsilon _1)+C_1 M(q_2,\varepsilon _1)\right)
    \nonumber
    \\
    &\hspace{2cm}-C_1 \varepsilon _1\cdot \varepsilon _2 M(q_1,q_2)\Big)+(p_1+p_2)^2 \Delta \mathfrak{C}_s,
\end{align}
from which we see that $C_1 =1$. We conclude that the color-kinematics duality agrees with the factorizations of the amplitude for the following values of the Wilson coefficients,
\begin{equation}
    C_1 = 1~,\hspace{0.3cm} D_1 = 0,
\end{equation}
providing further proof that imposing the color-kinematics duality comes at the cost of relinquishing freedom in the linear-in-spin Wilson coefficients \cite{Carrasco:2023wib, Carrasco:2023qgz}. We have found, however, that the at the classical level, and up to quadratic-in-spin order, this procedure puts no further constraints on Wilson coefficients.

\subsubsection{Gauge invariance}
Many, but not all of the free coefficients of the ansatz in \meqref{Eq:4-point linear ansatz}
are fixed on the two factorization channels above. The remaining freedom is fixed by imposing that the QED and QCD amplitudes, given in \meqref{Eq: 4-point amplitudes def qedym}, is gauge invariant. We impose gauge invariance by requiring the following, 
\begin{align}
    \mathcal{A}^{\text{QED}}(p_1,p_2,q_1,q_2) \big|_{\varepsilon_i \to q_i} = 0, \nonumber\\
    \mathcal{A}^{\text{QCD}}(p_1,p_2,q_1,q_2) \big|_{\varepsilon_i \to q_i} = 0.
\end{align}
We investigate gauge invariance  for the different orders in spin separately: the scalar terms are trivially gauge invariant, while imposing invariance on the spin-one terms constrain free ansatz coefficients. There are still some free coefficients in the numerator after imposing GI and factorization properties, but these are canceled out of the full amplitudes, and so we set them to zero in the final result. The final expression for the numerator function is then,
\begin{equation}
    n\left(~
\begin{gathered}
    \includegraphics[scale=1.0]{figures/spin_4p_t.pdf}
\end{gathered}
~\right)
=
\begin{gathered}
\vspace{1.5cm}\\
i\Big[4i (p_1 \cdot \varepsilon_2) (p_2 \cdot \varepsilon_1) 
-i\big(2( p_2\cdot q_2)+(p_1+p_2)^2\big) (\varepsilon_1 \cdot \varepsilon_2)\\
-2 (p_1 \cdot \varepsilon_2)  M(2 q_1+q_2, \varepsilon_1) 
+2 (\varepsilon_1 \cdot \varepsilon_2) M(q_1,q_2)  \\
+ (p_1+p_2)^2 M(\varepsilon_1,\varepsilon_2)  
+2((p_1+p_2)\cdot \varepsilon_1) M(q_1, \varepsilon_2) \\
-2  (p_2\cdot\varepsilon_2) M(q_2, \varepsilon_1)
+4(p_2 \cdot \varepsilon_1) M(q_2, \varepsilon_2) 
\Big] .
\end{gathered}
\end{equation}
The numerator of the other topology is then given by \meqref{Eq: 4-point Jacobi}, 
\begin{equation}
n\left(~
\begin{gathered}
  \includegraphics[scale=1.0]{figures/spin_4p_s.pdf}
\end{gathered}
~\right)
=
\begin{gathered}
\vspace{1cm}\\
2 \Big[2 ((p_1 \cdot \varepsilon_1)(p_2 \cdot \varepsilon_2)
- (p_1 \cdot\varepsilon_2) (p_2 \cdot \varepsilon_1))
+\left(p_2\cdot (q_2- q_1)\right)(\varepsilon_1 \cdot \varepsilon_2)
\Big]
\\2 i \Big[2 (p_{12} \cdot\varepsilon_1) M(q_1+q_2, \varepsilon_2) 
-2 (p_{12}\cdot \varepsilon_2)  M(q_1+q_2, \varepsilon_1)
\\
 +2  (\varepsilon_1 \cdot \varepsilon_2) M(q_1,q_2)
+(p_1+p_2)^2 M(\varepsilon_1, \varepsilon_2)
\Big].
    \end{gathered}
\end{equation}

\subsubsection{Quadratic-in-spin four-point amplitude}

Bootstrapping the quadratic-in-spin four-point amplitude follows the same logic as the linear-in-spin procedure presented above. In this case, however, we find that the quadratic Wilson coefficients $C_2$ and $D_2$ are \textit{not} constrained by the color-kinematics duality, but are left free. 
Same as with the linear-in-spin order, the appearance of products of Lorentz generators and the subsequent application of Lorentz algebra relations, may introduce quantum corrections, even when our seed has been tailored to be classical. In principle, products of Lorentz generators up to very high order may appear, and need to be reduced even up to linear-in-spin terms. However, we choose to keep no more than one Lorentz algebra reduction at four points (and two at five points), as we deem other terms too quantum to ever be relevant.
An interesting issue related to this is that there are a few free coefficients in the ansatz, which are not fixed by factorizations, colour-kinematics duality or gauge invariance. However, we find that they don't contribute at the classical level. 

Let us show explicitly the four-point Compton amplitude
\begin{equation}
    \mathcal{A}_{4}(p_1, p_2,q_1, q_2) = 
    \begin{gathered}
\includegraphics[scale=1.0]{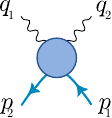}
\end{gathered},
\end{equation}
in Abelian gauge theory (QED), and up to quadratic-in-spin order. We 
recall here the relation for the classical limit of the amplitudes,
\begin{align}
    M^n(v_1,v'_1,\ldots,v_n,v'_n)
    =
    S(v_1,v_1')S(v_2,v_2')\ldots S(v_n,v_n'),
\end{align}
where we use $S(a,b)\equiv S^{\mu\nu}a_\mu b_{\nu}$, and $S^{\mu\nu}$ is the classical spin tensor. 
Furthermore, we consider the restoring of the $\hbar$ factors, following ref.  \cite{Maybee:2019jus}
\begin{align}
    q_i=\hbar\ q_i, \qquad S^{\mu\nu}=\hbar^{-1}S^{\mu\nu}.
\end{align}
The classical limit of the bootstrapped Compton amplitude then results in
\begin{align}
\mathcal{A}_{4,\text{cl.}}=\frac{m^2}{(p\cdot\kk_1)^2}(\omega_0+\omega_1+\omega_2),
\label{A4cl}
\end{align}
where we have defined $p\equiv p_1$. We express the amplitude in terms of the manifestly gauge-invariant field-strength functions $F_i^{\mu\nu} \equiv \varepsilon_i^\mu q_i^\nu - \varepsilon_i^\nu  q_i^\mu$. In terms of them, the spin-independent term is
\begin{align}
m^2\omega_0=2\,p\cdot F_1\cdot F_2\cdot p\,.
    \label{eq:s0_comp}
\end{align}
Similarly, evaluating the linear-in-spin part leads to
\begin{align}
\label{eq:s1_comp}
m\omega_1 =i(S \cdot F_1) \  \kk_{1}\cdot F_2\cdot p+i(S\cdot F_2) \  \kk_{2}\cdot F_1\cdot p+2i(p\cdot \kk_1)F_2\cdot S\cdot F_1,
\end{align}
while the quadratic-in-spin terms are 
\begin{align}
\label{eq:s2_comp}
m^2\omega_2=
&
m^2(\kk_1\cdot\kk_2)S\cdot F_1\  S\cdot F_2 /4 \\ \nonumber
&+C_2\ \Big\lbrace p\cdot F_1\cdot S \cdot S \cdot \kk_1\ p\cdot F_2 \cdot \kk_{1} \\ \nonumber
&+(p\cdot\kk_1)\big[ p\cdot F_1 \cdot F_2\cdot S\cdot S \cdot \kk_2+ 2p \cdot F_2\cdot S \cdot S\cdot F_1 \cdot \kk_{2}\big]\\ \nonumber 
&+(p\cdot\kk_1)^2F_1 \cdot S \cdot S \cdot F_2\Big\rbrace \quad+\quad (1\leftrightarrow 2).
\end{align}
In the previous two equations we use the notation
\begin{align}
T_1 \cdot T_2 \cdot \ldots \cdot T_n =  \eta_{\mu_1 \nu_1} \eta_{\mu_2 \nu_2} \ldots \eta_{\mu_n \nu_n} \eta_{\mu_0 \nu_0} T_1^{\nu_0 \mu_1} T_2^{\nu_1 \mu_2} \ldots T_n^{\nu_n \mu_0}, 
\end{align}
for any string containing only tensors $T_1,T_2,\ldots,T_n$.
This amplitude had previously been obtained using Feynman diagrams, with the relevant propagators and three- and four-point vertices derived from the field theory Lagrangian eq. (\ref{eq:LS1}).
The amplitude $\mathcal{A}_\text{4,cl}$ is therefore the $C_1=1$ and $D_1=0$ limit of the one in ref. \cite{Bern:2023ity}, which explicitly depend on both the $C_j$ and $D_j$ Wilson coefficients. Notably, at classical level the aforementioned limit also gets rid of any dependence on $D_2$.
Furthermore, in the black hole limit $C_2=1$, these are equivalent in four dimensions to the spin multipole coefficients $\omega_i$ of ref. \cite{DeAngelis:2023lvf},
which were obtained in the context of a gravity computation. We will come back to that result in the double copy section below. Before that, let us comment on the generalisation of the procedure to higher points.

\subsection{Bootstrapping five-point amplitudes}
The strategy for bootstrapping the five-point amplitude is the same as in the previous section --- we use the BCJ relations to determine a basis graph to be dressed with an ansatz for its kinematic numerator, which is then fixed on color-kinematic properties, factorization channels, and gauge invariance of the amplitude. The five-point amplitude is described in terms of three unique topologies, 
\begin{equation}\label{Eq: 5-point topologies}
\begin{aligned}
    \begin{gathered}
        \includegraphics[scale=1.0]{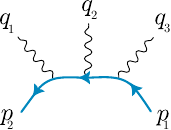}
    \end{gathered}
    \hspace{0.4cm}
    ,
    \hspace{0.4cm}
    \begin{gathered}
        \includegraphics[scale=1.0]{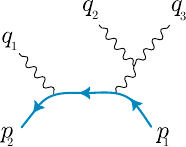}
    \end{gathered}
    \hspace{0.4cm}
    ,
    \hspace{0.4cm}
    \begin{gathered}
        \includegraphics[scale=1.0]{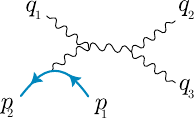}
    \end{gathered},
    \\
n_5^1(p_1,p_2,q_1,q_2,q_3)
\hspace{0.8cm}
    n_5^2(p_1,p_2,q_1,q_2,q_3)
\hspace{0.8cm}
    n_5^3(p_1,p_2,q_1,q_2,q_3)
\end{aligned}
\end{equation}
where the associated kinematic numerator $n_5^i$ is stated below each topology. Similarly to the four-point amplitude, the numerators of the topologies can be related using the following BCJ relations 
\begin{align}
        \begin{gathered}
        \includegraphics[scale=1.0]{figures/spin_5p_2.pdf}
    \end{gathered}
    \hspace{0.2cm}
    &=
        \hspace{0.3cm}
    \begin{gathered}
        \includegraphics[scale=1.0]{figures/spin_5p_1.pdf}
    \end{gathered}
        \hspace{0.4cm}
    -
        \hspace{0.4cm}
    \begin{gathered}
        \includegraphics[scale=1.0]{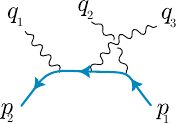}
    \end{gathered}
    \\
      \begin{gathered}
        \includegraphics[scale=1.0]{figures/spin_5p_3.pdf}
    \end{gathered}
    \hspace{0.4cm}
    &=
\hspace{0.3cm}
    \begin{gathered}
        \includegraphics[scale=1.0]{figures/spin_5p_2.pdf}
    \end{gathered}
    \hspace{0.2cm}
    -
    \hspace{0.3cm}
    \begin{gathered}
        \includegraphics[scale=1.0]{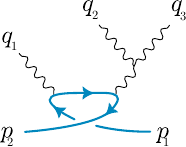}
    \end{gathered}
    ,
    \end{align}
which in terms of the numerator functions are expressed as 
\begin{align}
    n_5^2(p_1,p_2,q_1,q_2,q_3)
     &=    
     n_5^1(p_1,p_2,q_1,q_2,q_3)
         -
      n_5^1(p_1,p_2,q_1,q_3,q_2),\\
       n_5^3(p_1,p_2,q_1,q_3,q_2) 
     &= 
      n_5^2(p_1,p_2,q_1,q_3,q_2)
      -
       n_5^2(p_2,p_1,q_1,q_3,q_2).
\end{align}
From these equations we see that $n_5^1$ can be treated as the basis graph, and so we construct an ansatz for its numerator in a similar way that we did for $n_4^1$ in the previous section. The ansatz now consists of 1188 terms of the form,
\begin{equation}\label{Eq:5-point linear ansatz}
\begin{aligned}
\mathfrak{A}_1(p_1,p_2,q_1,q_2, q_3) =& ~\sum a_i \beta^L_i \beta^L_j (\varepsilon^s_1 \cdot \varepsilon^s_2)
+\sum b_i \beta_i^L \beta_j^M
+\sum c_i\beta_i^L \beta_j^L \beta_k^M
\\
&+\sum d_i \beta_i^L \beta_j^L \beta_k^L \beta_l^M,
\end{aligned}
\end{equation}
where the basis of Lorentz products is now,
\begin{multline}
    \beta^L = \big\{
    p_1\cdot p_2~,~p_1\cdot q_2~,~p_1\cdot q_3~,~p_1\cdot \varepsilon _1~,~p_1\cdot \varepsilon _2~,~p_1\cdot \varepsilon _3~,~m^2,
    \\
    p_2\cdot q_2~,~p_2\cdot q_3~,~
    p_2\cdot \varepsilon _1~,~
    p_2\cdot \varepsilon _2~,~p_2\cdot \varepsilon _3~,~q_2\cdot \varepsilon _3~,
    \\
    q_3\cdot \varepsilon _1~,~q_3\cdot \varepsilon _2~,~\varepsilon _1\cdot \varepsilon _2~,~\varepsilon _1\cdot \varepsilon _3~,~\varepsilon _2\cdot \varepsilon _3~,~\varepsilon^s_1 \cdot \varepsilon^s_2\big\} ,
\end{multline}
while the basis of Lorentz matrices is,
\begin{multline}
    \beta^M = 
\big\{
M(q_1,q_2)~,~M(q_1,q_3)~,~M(q_1,\varepsilon _1)~,~M(q_1,\varepsilon _2)~,~M(q_1,\varepsilon _3),\\
M(q_2,q_3)~,~
M(q_2,\varepsilon _1)~,~M(q_2,\varepsilon _2)~,~
M(q_2,\varepsilon _3)~,~M(q_3,\varepsilon _1),\\
M(q_3,\varepsilon _2)~,~
M(q_3,\varepsilon _3)~,~M(\varepsilon _1,\varepsilon _2)~,~M(\varepsilon _1,\varepsilon _3)~,~M(\varepsilon _2,\varepsilon _3)
\big\}.
\end{multline}
The coefficients of the ansatz are fixed by imposing that the numerators obey the three maximal cut factorizations of the topologies in \meqref{Eq: 5-point topologies}. Unsurprisingly, we find the same constrains as with the result from the previous section --- namely that having color-dual numerators requires the Wilson coefficients to be $C_1 = 1$ and $D_1=0$.

Similarly to the four-point amplitude, we fix a subset of the coefficients by imposing gauge invariance of the QED and QCD amplitudes. The scalar part of the amplitude is then completely fixed. For the linear-in-spin part of the amplitude there are still free coefficients after imposing gauge invariance. However, they only appear in terms which are suppressed in the classical limit and therefore don't impact the classical result.  

Covariant results for the five-point scattering amplitudes were first obtained in ref. \cite{Bjerrum-Bohr:2023jau}, while a direct comparison is nontrivial due to the spin structures being non-independent, we have resorted to numerical checks to confirm the equivalence between our results. We expect colour-kinematics to hold at the quadratic-in-spin level also for the five-point amplitude. And so we could proceed analogously, with the only difference of the ansatz being larger, and the procedure becoming more time consuming.

We include results for amplitudes from scalar to quadratic-in-spin order for QED, QCD and GR at three-point and four-point levels respectively in  the ancillary files {\tt ThreePoint.m} and {\tt FourPoint.m}. In the case of five points, we include the scalar and the linear-in-spin result for the three theories in the file {\tt FivePoint.m}.

\section{Gravity Amplitudes and the Double copy}
\label{sec::Gravity}
In the previous section we bootstrapped kinematic numerators for a gauge theory such that the amplitude can be expressed in terms of the numerators and color factors of cubic graphs as
\begin{align} \mathcal{A}=\sum_{i\in \textrm{cubic}} \frac{n_i c_i}{d_i}.
\end{align}
With the gauge theory amplitudes in hand, we can write the classical limit of our gravity amplitudes by using the BCJ double copy relations
\begin{align} \mathcal{M}=\sum_{i\in \textrm{cubic}} \frac{n_i \tilde{n}_i}{d_i}.
\end{align}
Let us first take an asymmetric double copy where the first set of numerators come from the spinning amplitude, while the second set comes from the scalar. Generally, the double copy of gluons will lead not only to gravitons, but also a massless scalar state (dilaton) and an antisymmetric tensor (axion). For tree-level amplitudes with more than one massive state where dilatons and axions can propagate between massive lines, and one-loop amplitudes, these states have to be removed. Progress in this direction was recently made by one of the authors (IV-H) in ref. \cite{Johansson:2025grx}. In the present case however, all massless states can be controlled by projecting the amplitudes on the polarization of the graviton.
We have verified that this amplitude can be equivalently expressed as the KLT-like relation
\begin{align}
\label{M4DCS0}
\mathcal{M}_{4,\text{cl.}}=\frac{(2p\cdot q_1)^2}{2q_1\cdot q_2}
\big(\mathcal{A}_{4,\text{cl.}}\big)
\big(\mathcal{A}_{4,\text{cl.}}|_{S^0}\big),
\end{align}
where $\mathcal{A}_{4,\text{cl.}}$ is given in eq. (\ref{A4cl}), and $\mathcal{A}_{4,\text{cl.}}|_{S^0}$ is its spinless limit. Since the amplitude eq. (\ref{A4cl}) is quadratic-in-spin order, it produces the gravity amplitude at that same order, which reproduces the one obtained in \cite{Bern:2022kto}.
The match for both the scalar and the linear-in-spin parts holds also at the quantum level. Starting at quadratic-in-spin, the double copy we produce only matches in the classical limit. 

By building the double copy of a spinning particle with a scalar one, the highest order we can get in gravity  is the same as the one we have in gauge theory. 
However, we are also interested in cubic and quartic orders in spin. Furthermore it has been shown that one can obtain via double copy amplitudes with quantum spin \cite{Johansson:2019dnu, Bautista:2019evw}, from which these orders in the multipole expansion can be obtained.
To do this, however, it's necessary to consider a double copy between both spinning particles. 
As a practical matter, this requires relations for Lorentz generators of different representations. We will consider our generators to be in the representation $(s+1,s+1)$ where the two entries $(d_L,d_R)$ are the dimensions of the two $SU(2)$ in the Lorentz group $SU(2)\times SU(2)\simeq SO(3,1)$. This problem was already considered in ref. \cite{Bern:2020buy}. To double copy two gauge amplitudes, one with $s=s_L$ and the other with $s=s_R$, 
one needs the projection of the product $(s_L+1, s_L+1)\times (s_R+1, s_R+1)$ 
onto $(s+1, s+1)$ with $s=s_L+s_R$. 
The relevant projection results in \cite{Bern:2020buy}
\begin{align}
\label{general_product}
&(M^{\mu_1 \nu_1}\dots M^{\mu_n \nu_n}) {}_{a({s_L})} {}^{b({s_L})}  \otimes (M^{\rho_1 \sigma_1}\dots M^{\rho_m \sigma_m})_{a({s_R})} {}^{b({s_R})} \Big| 
\\ \nn 
&\qquad\qquad\qquad\qquad
= {\cal C}(n,m, s_L, s_R)
(M^{\mu_1 \nu_1}\dots M^{\mu_n \nu_n} M^{\rho_1 \sigma_1}\dots M^{\rho_m \sigma_m})_{a({s})} {}^{b({s})} \,,
\end{align}
where 
\begin{equation}
{\cal C}(n,m, s_L, s_R) = \frac{s_L!}{(s_L-n)!}\frac{s_R!}{(s_R-m)!}\frac{(s-n-m)!}{s!} \,.
\end{equation}
Using this, we pick representations for the theories with $s_L=s_R=2$. The Clebsch-Gordan coefficients are
\begin{align}
    \nonumber
    {\cal C}(0,0) &= 1,\qquad 
    {\cal C}(1,0) =
    {\cal C}(0,1) = 1/2,\qquad
    {\cal C}(1,1) = 1/3,\\
    {\cal C}(2,0) &=
    {\cal C}(0,2) =
    {\cal C}(2,1) =
    {\cal C}(1,2) =
    {\cal C}(2,2) = 1/6,
\end{align}
where we use the shorthand  ${\cal C}(i,j)\equiv {\cal C}(i,j,2,2)$. Upon use of this decomposition, we obtain a double copy amplitude up to $S^4$. In the black hole limit $C_2=1$ this results in 
\begin{equation}
  \label{eq:Kerr_compton}
   \mathcal{M}_{4,\rm cl,\rm BH} =\frac{\kappa^2 m^4}{ (2q_1 \cdot q_2) (2p_1 \cdot q_1)^2} \Big(\omega_0^2 + \omega_0 \omega_1 + \omega_0 \omega_2 + \frac{\omega_1 \, \omega_2}{3} + \frac{\omega_2^2}{6}\Big)\,,
\end{equation}
where $\omega_i$ are defined in eqs. (\ref{eq:s0_comp})-(\ref{eq:s2_comp}). Note that to express it in this form, we have used the relation $2\omega_0\omega_2=\omega_1^2$, which only holds in the black hole limit.
This compact form for the covariant Compton amplitude in gravity (which matches the solution of the Teukolsky equation)  was first found in Ref. \cite{DeAngelis:2023lvf}\footnote{
It was also shown in ref. \cite{DeAngelis:2023lvf}, that eq. \eqref{eq:Kerr_compton} can be equivalently written as the exponential
\begin{equation}
\nonumber 
  \label{eq:Kerr_exp}
\mathcal{M}_{4,\rm cl,\rm BH} = \frac{\kappa^2 m^4\omega_0^2}{ (2q_1 \cdot q_2) (2p_1 \cdot q_1)^2} \exp \left( \frac{\omega_1}{\omega_0} \right) + \mathcal{O}(S^5) \, .
\end{equation}
This form, however, features a spurious pole in $\omega_0$ starting at $\mathcal{O}(S^3)$.}. 

\subsection{The case for cubic-in-spin  color-kinematics}
One important feature of the double-copied amplitude in eq. (\ref{M4DCS0}) was that it reproduced the result also for a generic object (e.g. neutron star). However, eq. (\ref{eq:Kerr_compton}) only contains the information for the black hole limit. This raises the question: can we obtain the generic object through a double-copy procedure beyond quadratic-in-spin order? The amplitude has been computed before. The first covariant form of the amplitude was obtained and used (though unpublished) in ref. \cite{Bern:2022kto}, using Feynman rules derived from the Lagrangian included here in the appendix \ref{app:Lag}. A more compact (though still containing a few hundred terms) version of the amplitude is given in \cite{Ben-Shahar:2023djm}. Direct comparisons of covariant amplitudes are not straightforward due to relations between spin structures (stemming from Gram determinants) obscure the equivalence of different results. For this reason, we find it useful to consider helicity amplitudes instead of the covariant results. For concreteness, we focus here on the $(1^{-},2^{+})$ helicity configuration, for which the amplitude takes the form
\begin{align}
\label{comptonNSS3}
\mathcal{M}^{-+}_{\text{4,cl.}}&=\frac{iy^4}{2stu}\,\Big\{(\qt_{12}\cdot a)+\frac{1}{2}(\qt_{12}\cdot a)^2
+\frac{1}{6}(\qt_{12}\cdot a)^3
\\
\nn
&
+\frac{1}{2}\widetilde{C}_{\text{2,GR}}\left((\qt_1\cdot a)^2+(\qt_2\cdot a)^2\right)+\frac{1}{6}\widetilde{C}_{\text{3,GR}}\left((\qt_1\cdot a)^3+(\qt_2\cdot a)^3\right)\\
\nn 
&
+(\widetilde{C}^2_{\text{2,GR}}-\widetilde{H}^2_{2})(\qt_1\cdot a)(\qt_2\cdot a)(\omega_1\cdot a)-\frac{\widetilde{C}_{\text{2,GR}}}{2}(\qt_1\cdot a)(\qt_2\cdot a)(\qt_{12}\cdot a)
\Big\},
\end{align}
where we have used the definitions
\begin{align}
\nonumber 
\omega_\mu \equiv \frac{1}{2}\langle 1|&\sigma_\mu|2],\qquad 
y\equiv2 p_1\cdot \omega,\qquad
\omega_i^\mu\equiv \frac{2p_1\cdot q_i}{y}\omega^\mu, \\
    \qt_i^\mu &\equiv q_i^\mu-\omega_i^\mu,\qquad
    \qt_{12}^\mu\equiv \qt_{1}^\mu-\qt_{2}^\mu.
\end{align}
The appearing combinations of Wilson coefficients are
\begin{align}
\widetilde{C}_{\text{2,GR}}\equiv C_{ES^2}+H_2-1,
\qquad
\widetilde{C}_{\text{3,GR}}\equiv C_{BS^3}+H_3-1,\qquad
\widetilde{H}_{2}\equiv H_2-1,
\end{align}
in terms of which the black hole limit is 
$\widetilde{C}_{\text{2,GR}}=\widetilde{C}_{\text{3,GR}}=\widetilde{H}_{2}=0$. The Wilson coefficient $\widetilde{H}_2$ is associated with a Generic Compact Object (GCO), while the coefficients $\widetilde{C}_{\text{2,GR}}$ and $\widetilde{C}_{\text{3,GR}}$ corresponds to those of a Conventional Compact Object (CCO), as defined in ref. \cite{Alaverdian:2025jtw} (although the naming of the Wilson coefficients is different there). 
The amplitude for the CCO (meaning for $\widetilde{H}_2=0$) has been also obtained using a classical worldline in refs. \cite{Saketh:2022wap}, and with recursion relations in ref. in refs. \cite{Haddad:2023ylx}. 
Let us circle back to the question of whether the amplitude  can be obtained through double copy. 
Since even the amplitude for the CCO has the Wilson coefficient $C_{\text{3,GR}}$, our numerator doesn't have enough free parameters as it is, so matching this amplitude would rely on finding a colour-kinematic dual gauge amplitude with a corresponding Wilson coefficient at cubic-in-spin level.
The question of the feasibility of our bootstrap procedure at cubic-in-spin order is central in our study, but we have been so far unable to fix numerators with this property in our current framework. To understand where color-kinematics duality breaks down for cubic-in-spin order, it is again useful to look at the factorization channels and the discrepancy function, which is given in \meqref{Eq: pseudo-Jacobi}. In the cubic-in-spin case, the massless factorization is, 
\begin{equation}
\begin{aligned}
\begin{gathered}
\includegraphics[scale=1.0]{figures/spin_4p_s_cut.pdf}
\end{gathered}
=& 2  \frac{D_3}{m^2}
\Big[
(\varepsilon_1\cdot \varepsilon_2) M^3(\xi,k_{12},\xi,k_{12},k_1,k_2)
\\
&
-
(k_{34}\cdot \varepsilon_2) M^3(\xi,k_{12},\xi,k_{12},k_{12},\varepsilon_1)
+(k_{34}\cdot \varepsilon_1)
M^3(\xi,k_{12},\xi,k_{12},k_{12},\varepsilon_2)
\Big]. 
 \end{aligned}
\end{equation}
The massive factorization can again be split into the two terms $\mathfrak{C}_{\beta_1}$ and $\mathfrak{C}_{\beta_2}$, where the bases $\beta_1$ and $\beta_2$ now consist of $M^3$'s that show up in both factorizations, and only the massive factorization, respectively. The difference between the two massive factorizations is again, 
\begin{equation}
     \begin{gathered}
\includegraphics[scale=1.0]{figures/spin_4p_t_cut.pdf}
\end{gathered}
-
\begin{gathered}
\includegraphics[scale=1.0]{figures/spin_4p_u_cut.pdf}
\end{gathered}
= \Delta \mathfrak{C}_{\beta_1} + \Delta \mathfrak{C}_{\beta_2}, 
\end{equation}
where we take a closer look at $\mathfrak{C}_{\beta_1}$, 
\begin{equation}
\begin{aligned}
\Delta \mathfrak{C}_{\beta_1} 
=&   \hspace{0.5cm}
2\frac{  C_3 m^2 \varepsilon _1\cdot \varepsilon _2+F_3 \left(p_1\cdot \varepsilon _2 p_2\cdot \varepsilon _1+p_1\cdot \varepsilon _1 p_2\cdot \varepsilon _2\right)}{m^6} M^3(q_1,q_2,q_1,q_2,q_1,q_2)
\\
&+2\frac{ C_3 }{m^4}\left(p_{12}\cdot \varepsilon _1 M^3(q_1,q_2,q_1,q_2,q_1,\varepsilon _2)-p_{12}\cdot \varepsilon _2 M^3(q_1,q_2,q_1,q_2,q_2,\varepsilon _1)\right)
\\
&-\frac{2 i }{m^4}\Big[
M^3(q_1,q_2,q_1,q_2,q_2,\varepsilon _2) \left(D_2^2 p_1\cdot \varepsilon _1-\left(D_2-1\right) F_2 p_2\cdot \varepsilon _1\right)
\\
& \hspace{1cm} -M^3(q_1,q_2,q_1,q_2,q_1,\varepsilon _1) \left(D_2^2 p_1\cdot \varepsilon _2-\left(D_2-1\right) F_2 p_2\cdot \varepsilon _2\right)
\Big]
\\
&-4 i\frac{ \left(D_2-2\right) D_2 }{m^2} M^3(q_1,q_2,q_1,\varepsilon _1,q_2,\varepsilon _2).
\end{aligned}
\end{equation}
For the Jacobi identity to be satisfied, this part of the discrepancy function must vanish up to inverse propagators, which could only happen if all Wilson coefficients are set to zero. This means indicating that no color-kinematics dual numerators can be found for this graphs, which are compatible with the cubic-in-spin three-point amplitudes we have considered here.

\section{Discussion}
\label{sec::discussion}
In this paper, we have established a systematic procedure to generate numerators for Compton-like amplitudes in gauge theories, exploiting colour-kinematics duality. We have used graphs to determine the Compton amplitude in the framework of the theory of arbitrary integer spin introduced in \cite{Bern:2020buy}. These amplitudes describe arbitrary spinning bodies (like a neutron star in gravity) up to quadratic-in-spin order at four points, and five point linear-in-spin.
Our procedure shares some traits with the program of refs. \cite{Bjerrum-Bohr:2023jau,Bjerrum-Bohr:2023iey,Bjerrum-Bohr:2024fbt}, which obtains compact high-order in spin expressions for HEFT amplitudes building on the construction of color-kinematic dual numerators of ref. \cite{Brandhuber:2021bsf}. Our focus was however on the exploitation of colour-kinematics for the constraint of a most generic Compton amplitude, as well as on the exploration of the validity of colour-kinematics in our setup. Indeed, using a discrepancy function, we have studied the way colour kinematics constrains the Wilson coefficients in our theory. We find, as expected that this duality selects the gyromagnetic ratio $C_1=1$. This behaviour was first observed by Holstein in ref. \cite{Holstein:2006pq}, and later understood as a KLT relation in ref. \cite{Bjerrum-Bohr:2013bxa}. The equivalence between double copy and gyromagnetic ratio $g=2$ was thoroughly studied for (quantum) lower spins in ref. \cite{Bautista:2019evw}. 

The quadratic-in-spin numerator can be double copied with itself to obtain quartic-in-spin results. This is done along the lines of the covariant spin multipole double copy from refs. \cite{Bautista:2019evw,Bautista:2022wjf}, except that our setup makes it simpler to work with representations of the group $SU(2)\times SU(2)$ and their products, as introduced in ref. \cite{Bern:2020buy}. The question of understanding the equivalence between the Clebsch-Gordan decomposition in both setups is an interesting one, but we defer it to later work. Upon building this double copy, we reproduce the result for the black hole after setting the Wilson coefficient $C_2=1$. 
An attempt to either go to higher than quartic order in spin, or to describe a generic body is prevented by not finding colour-kinematics dual numerators at cubic-in-spin order. 
Indeed, the discrepancy function shows that this is incompatible with the ansatz we take as  starting point.
This process seems to be related to a previous finding by Ochirov and Johansson in cite \cite{Johansson:2019dnu}, where an inconsistency for the double copy at that order is pointed out. In that case this manifests as (not removable) spurious singularities in the factorisation in terms of one spinning and one scalar single copies. 
There are, however, important physical scenarios with higher spin where certain notions of double copy can be implemented, for example string theory. Another instance, closer to the scope of this paper, is that we know that the Kerr black hole can be understood as a double copy working to all orders in spin \cite{Monteiro:2014cda}.  
In future work, it would be very interesting to investigate the relation between the graphical methods based on color-kinematics employed here, and the scenarios mentioned above. In principle, a more general ansatz would be in order to allow for color-kinematics duality in those settings. However, understanding the Kerr black hole as a double copy is already a subtle issue, because so far this has only been treated in an abelian way.

Finally, we can extend our construction to higher-points systems, and we expect that our method can be used in the production of observables for the description of the two-body dynamics (or radiation) of a binary of compact objects.

\section*{Acknowledgements}
We thank Fabian Bautista, Maor Ben-Shahar, John Joseph Carrasco, Gang Chen, Stefano De Angelis, Riccardo Gonzo, Kays Haddad, Henrik Johansson, Marcos Skowronek, and Fei Teng for useful discussions.
A.~L. is supported by funds from the European Union’s Horizon 2020 research and innovation program under the Marie Sklodowska-Curie grant agreement No.~847523 ‘INTERACTIONS’, and by the Villum Foundation Grant No. VIL37766. I. V. H. is supported by the Knut and Alice Wallenberg Foundation under grants KAW 2018.0116 and KAW 2018.0162.

\appendix

\section{Appendix}

\subsection{The gravity Lagrangian}
\label{app:Lag}
The minimal Lagrangian in gravity has the standard two-derivative kinetic terms
\begin{align}
{\cal L}_\text{GR}=-R,\qquad {\cal L}_\text{min} =\frac{1}{2} \eta^{ab} \nabla_a\phi_s\nabla_b\phi_{s}
- \frac{1}{2}m^2\phi_s\phi_{s} \ ,
\label{Ls}
\end{align}
where we use tangent-space indices.
We take the higher-spin field $\phi_s$ to be in a real representation of the Lorentz group. In the non-minimal Lagrangian, we consider two classes of linear-in-curvature operators
\begin{align}
    \mathcal{L}_\text{non-min}=\mathcal{L}_C+\mathcal{L}_H.
\end{align}
The first two linear-in-curvature operators for the first class are
\begin{align}
\label{Lnonmin}
{\cal L}_C  =& -\frac{1}{2!}
\frac{C_{\textrm{ES}^{2}}}{m^{2}}  R_{a f_1 b f_2 }
\nabla^a{\phi}_s\,  \S^{(f_1}\S^{f_{2})} \nabla^b \phi_s
\\
&\null + \frac{1}{3!}
\frac{C_{\textrm{BS}^{3}}}{m^{3}}  \nabla_{f_3} \widetilde{R}_{(a|f_1|b)f_2}
\nabla^a{\phi}_s  \S^{(f_1} \S^{f_2} \S^{f_{3})}  \nabla^b\phi_s
\,,
\nonumber
\end{align}
where $\S^{a}\equiv \frac{-i}{2m}\epsilon^{a b cd}M_{cd}\nabla_{b}$ and $\widetilde{R}_{abcd}\equiv\frac{1}{2}\epsilon_{abij}R^{ij}{}_{cd}$ is the dual Riemann tensor.
The operators in Eq.~\eqref{Lnonmin} are in one-to-one correspondence with the first non-minimal operators 
in Ref.~\cite{Levi:2015msa}. The first terms in the second family of linear-in-curvature operators we include are given by
\begin{align}\label{Hs}
\mathcal{L}_{H}  =&  \frac{1}{3\cdot 2!}H_{2}R^{(a}{}_{f_1}{}^{b)}{}_{f_2}\phi_sM_{a}{}^{(f_1}M_{b}{}^{f_2)}\phi_s\nonumber\\
&-\frac{1}{2\cdot 3!}\frac{H_{3}}{m}\nabla_{f_3}\widetilde{R}^{(a}{}_{f_1}{}^{b)}{}_{f_2} \phi_sM_{a}{}^{(f_1}M_{b}{}^{f_2}\S^{f_3)}\phi_s\, .
\end{align}
The normalization is chosen such that the three-point amplitudes depend only on
$C_{2}\!\equiv\! C_{\textrm{ES}^{2}}{+}H_{2}$ and $C_{3}\!\equiv\! C_{\textrm{BS}^{3}}{+}H_{3}$.
Comparison with Ref.~\cite{Vines:2017hyw} fixes $C_3=C_2=1$ for a Kerr black hole. 
Both  $\mathcal{L}_C$ and $\mathcal{L}_{H}$ contribute differently to the classical gravitational Compton amplitude, so both need to be included in the EFT.

\bibliographystyle{JHEP}
\bibliography{mainbib}

\end{document}